\documentclass[aps,prd,twocolumn,groupedaddress,nofootinbib]{revtex4-1}

\usepackage{dsfont}
\usepackage{natbib}
\usepackage{amssymb}
\usepackage{graphicx}
\usepackage{amsmath}
\usepackage{url}
\usepackage{epstopdf}
\usepackage{color}
\usepackage{enumerate}

\newcommand{\ket}[1]{|#1\rangle}
\newcommand{\braket}[2]{\langle #1|#2\rangle}

\newcommand{\Hp}{\mathcal{H}_{\text{p}}}

\newcommand{\Mp}{M_\text{Pl}}
\newcommand{\M}{M_\star}
\newcommand{\Ntot}{N_\text{tot}}
\newcommand{\di}{\partial}
\newcommand{\efolds}{N}

\newcommand{\gauge}{{t=f(a),\mathcal{H}=0}}

\def\p {\partial}

\def\be {\begin{equation}}
                                                                                
\def\ee  {\end{equation}}
                                                                                
\def\bea {\begin{eqnarray}}
                                                                                
\def\eea {\end{eqnarray}}
                                                                                
\def\nn {\nonumber}

\begin{document}

\title{Polymer inflation}

\author{Syed Moeez Hassan}
\email{shassan@unb.ca}
\affiliation{Department of Mathematics and Statistics, University of New Brunswick, Fredericton, NB, Canada E3B 5A3} 

\author{Viqar Husain}
\email{vhusain@unb.ca}
\affiliation{Department of Mathematics and Statistics, University of New Brunswick, Fredericton, NB, Canada E3B 5A3} 

\author{Sanjeev S.\ Seahra}
\email{sseahra@unb.ca}
\affiliation{Department of Mathematics and Statistics, University of New Brunswick, Fredericton, NB, Canada E3B 5A3} 

\date{\today}

\begin{abstract}

We consider the semi-classical dynamics of a free massive scalar field in a homogeneous and isotropic cosmological spacetime.  The scalar field is quantized using the polymer quantization method assuming that it is described by a gaussian coherent state.  For quadratic potentials, the semi-classical equations of motion yield a universe that has an early ``polymer inflation" phase which is generic and almost exactly de Sitter, followed by a epoch of slow-roll inflation.  We compute polymer corrections to the slow roll formalism, and discuss the probability of inflation in this model using a physical Hamiltonian arising from time gauge fixing.  We also show how in this model, it is possible to obtain a significant amount of slow-roll inflation from sub-Planckain initial data, hence circumventing some of the criticisms of standard scenarios.  These results show the extent to which a quantum gravity motivated quantization method affects early universe dynamics.

\end{abstract}

\maketitle

\section{Introduction}

The standard model of inflationary cosmology is the coupled dynamics of gravity and matter (usually a scalar field) with the assumption of homogeneity and isotropy. The Large Scale Structure of our universe is described by the Friedmann-Robertson-Walker (FRW) metric, together with the quantum fluctuations of the metric and matter on this background. It is the quantum fluctuations, made real by inflation,  that are considered the basis for structure formation. The quantization method is the usual standard Schr\"{o}dinger/Fock quantization, where the Hilbert space  is chosen as the space of square integrable functions on the spatial manifold. (For a recent review see for example \cite{Baumann:2009ds}.)

A question of much interest is whether and how quantum gravity might affect the standard model of cosmology. This has been studied from various points of view including string theory \cite{McAllister:2007bg,Brandenberger:2011et}, non-commutative geometry \cite{Chu:2000ww,Lizzi:2002ib,Brandenberger:2002nq}, loop quantum gravity (LQG) \cite{Bojowald:2008zzb}, and alternative quantization methods that incorporate a fundamental length scale. 
Although the regime in which inflation occurs is presently considered  to be far from the Planck scale, in the absence of a complete quantum theory of gravity it is only through such models that one might see the effects of quantum gravity at lower energy scales \cite{Shankaranarayanan:2002ax,Hassan:2002qk}. 

Our purpose here is to study the effect of an alternative quantization method called polymer quantization. This is motivated from LQG but is different from it, sharing only the idea that an alternative set of classical variables is used as a basis for quantization in a Hilbert space distinct from that used in standard quantum theory.   The method itself is  general in that it may  be applied to any classical theory. A basic feature is that it introduces a length scale in addition to $\hbar$ into the quantum theory at the outset; standard quantization results appear in a well defined limit.

Polymer quantization was first introduced in the quantum mechanics setting \cite{Ashtekar:2002sn, Halvorson}, and applied to the scalar field at the kinematical level in \cite{Ashtekar:2002vh}.  A related method using different variables was  applied to scalar field theory dynamics on flat  spacetime \cite{Hossain:2009vd,Hossain:2010eb,Husain:2010gb}, and to other quantum mechanical systems  \cite{Husain:2007bj, Kunstatter:2008qx,Kunstatter:2009ua,Hossain:2010wy}. The Schr\"{o}dinger limit of polymer quantization has also been discussed in detail \cite{Fredenhagen:2006wp,Corichi:2007tf}.

We apply this quantization method to the scalar field in the context of inflationary cosmology.  In \cite{Hossain:2009ru},  two of the present authors studied the effects of a polymer quantized massless scalar field in a Friedmann-Robertson-Walker (FRW) background. It was demonstrated that: (i) at early times, there is a de-Sitter like inflationary phase that solves the horizon and flatness problems while successfully avoiding the big bang singularity; and (ii) the universe dynamically emerges from inflation at late times, where polymer quantization effects become small and classical results are recovered.  These results were obtained assuming vanishing scalar potential and zero cosmological constant.  

The key difference between the polymer model studied in \cite{Hossain:2009ru} and the conventional picture of inflation is that, in the former, the inflationary epoch extends infinitely far into the past and is characterized by a (virtually) constant Hubble factor.  The number of e-folds of this polymer-driven inflation is therefore infinite, so the requirement that we need $\gtrsim 60$ e-folds of accelerated expansion to explain the homogeneity and apparent flatness of the universe is trivially satisfied.

However, it is widely believed that inflation is also responsible for producing primordial perturbations with a spectrum consistent with observations.  The latest results from cosmic microwave background (CMB) experiments \cite{Ade:2013ktc} suggest that the power spectrum of these perturbations is $\propto k^{n_{s}-1}$ with $n_{s} \sim 0.9675$.  To obtain $n_{s} \ne 1$, we require that the Hubble factor (evaluated when perturbative modes exit the Hubble horizon) to vary slowly during inflation.  It is fairly easy to see that the Hubble variation in the polymer model of Ref.\ \cite{Hossain:2009ru} is far too small to be consistent with the observed CMB.

In order to recover the period of ``slow-roll'' inflation favoured by observations, we extend the model of \cite{Hossain:2009ru} to include a nonzero scalar potential.  We study this model  in the context  of a fixed time gauge, where the Hamiltonian constraint is solved strongly to yield a physical Hamiltonian that describes the evolution of the scalar field phase space variables.  As described in detail below, such a model has phases of both polymer and slow roll inflation.  We discuss implications for the probability of inflation, quantify the amount of e-foldings that can be obtained, and comment on how sub-Planckian initial conditions can lead to significant slow-roll in this model.

The broader context of this type of study comes from LQG: if at a fundamental level both gravity and matter are polymer quantized, then there will be a lower energy regime where the effects of polymer quantization of matter  filter down to an emergent semiclassical level. It is therefore of some interest to see if this could lead to signatures for cosmology. 

In Section \ref{sec:Hamiltonian cosmology} we review Hamiltonian cosmology from a  canonical perspective, and in particular introduce the ``e-fold time" gauge which is used in subsequent sections. Section \ref{sec:polymer} is a review of polymer quantization of a scalar field, followed in Section \ref{sec:EOMs} by a numerical and analytic study of the semiclassical evolution equations, and the effects of polymer quantization on the slow roll parameters.  Section \ref{sec:e-folds} is a discussion of the probability of inflation, followed by a summary section.  

\section{Hamiltonian cosmology}\label{sec:Hamiltonian cosmology}

We begin with the Arnowitt-Deser-Misner (ADM) canonical formulation of  general relativity coupled to a scalar field. The phase space variables are $(q_{ab},\pi^{ab})$ and $(\phi, p_\phi)$ and the action is 
\be
 S= \int d^3x\  dt  \ \left(\pi^{ab} \dot{q}_{ab}  + p_\phi \dot{\phi} - {\cal N}{\cal H} - \mathcal{N}^a {\cal C}_a   \right) , 
\ee
where
\begin{subequations} 
\bea
{\cal H} &=& \frac{2}{\Mp^{2}}  \frac{ \pi^{ab}\pi_{ab} - \frac{1}{2} \pi^2 }{\sqrt{q}}  - \frac{\Mp^{2}}{2} \sqrt{q} \, {}^{(3)} \!R\nn\\
   && +   \left[  \frac{p_\phi^2}{2\sqrt{q}} +  \frac{1}{2}  \sqrt{q}q^{ab} \p_a\phi \p_b \phi   + V(\phi) \right],\\
{\cal C}_a &=&  -D_b \pi^b_{\ a} +  p_\phi \p_a\phi,
\eea
\end{subequations}
Here, $\Mp^{2} = 1/8\pi G$ is the reduced Planck mass, $\mathcal{N}$ is the lapse, $\mathcal{N}^{a}$ is the shift, $q_{ab}$ is the spatial 3-metric with covariant derivative $D_{a}$, and $V(\phi)$ is the scalar field potential.

Reduction to homogeneity and isotropy is obtained by the parametrization
\be
q_{ab} = a^2(t) e_{ab}, \quad \pi^{ab} = \frac{p_a}{6a(t) } e^{ab}, 
\ee
where $e_{ab}$ is the flat Euclidean metric.  This gives the reduced action
\begin{subequations}
\bea
S_\text{R} &=&V_0 \int dt \left(p_a\dot{a} +p_\phi \dot{\phi} - {\cal N} {\cal H} \right),\\
{\cal H} &=& - \frac{p_a^2}{12 a\Mp^{2}} + \frac{p_\phi^2}{2a^3} + a^3 V(\phi),
\eea
\end{subequations}
where $V_0=\int d^3x$ is a fiducial volume.
 
Notice that the reduced action is invariant under the rescalings
\begin{equation}
	(a,p_{a},\phi,p_{\phi},V_{0}) \mapsto \left( \frac{a}{\kappa^{1/3}},\frac{p_{a}}{\kappa^{2/3}},\phi, \frac{p_{\phi}}{\kappa}, \kappa V_{0} \right). \label{rescaling}
\end{equation}
This symmetry is due to the invariance of the spatially flat FRW metric under spatial dilations:
\begin{equation}\label{eq:re-scalings}
	(a,x,y,z) \mapsto (\kappa^{-1/3}a,\kappa^{1/3}x,\kappa^{1/3}y,\kappa^{1/3}z).
\end{equation}

 Cosmological observables of interest may be easily written in terms of phase space variables. The Hubble parameter is 
 \be
 H = \frac{\dot{a}}{{\cal N}a} = \frac{\{a, {\cal N}{\cal H} \}}{{\cal N}a}  =  \frac{\{a, {\cal H} \}}{a} = -\frac{p_a}{6a^2 \Mp^{2}}.
 \ee
Note the for an expanding universe $H>0$ and hence $p_{a}<0$.  In terms of the Hubble parameter, the Hamiltonian constraint assumes the familiar form of the Friedmann equation
 \be\label{Feqn}
 H^{2} = \frac{\rho_{\phi}}{3\Mp^{2}},
 \ee
 where the scalar field density $\rho_{\phi}$ is given in terms of the scalar field Hamiltonian density $\mathcal{H}_{\phi}$ as follows:
 \begin{equation}
 	\rho_{\phi} \equiv \frac{\mathcal{H}_{\phi}}{a^{3}} \equiv \frac{p_\phi^2}{2a^6} + V(\phi).
 \end{equation}
 Other quantities of cosmological interest can also be expressed in this way.  For example, the Hubble slow roll parameters are
 \begin{subequations}
 \bea
 \varepsilon_H &=& - \frac{\dot{H}}{\mathcal{N} H^{2}}  =  -\frac{ \{ H, {\cal H } \}}{H^2}  = \frac{3}{2} \left( \frac{{\cal L}_\phi}{{\cal H}_\phi}    +1   \right),\\
 \eta_H &=& \varepsilon_{H} - \frac{\dot{\varepsilon_{H}}}{2\mathcal{N}H\gamma_{H}} = - \frac{\{\{H, {\cal H}\}, {\cal H} \} }   { \{H^2, {\cal H} \} },
 \eea
 \end{subequations} 
 where $\displaystyle {\cal L}_\phi =  \frac{p_\phi^2}{2a^3} - a^3 V(\phi) $.

 The dynamics of the system can be obtained by solving Hamilton equations
\begin{equation}\label{eq:EOMs from constraint}
	\frac{d}{dt} ({a},{p}_{a},{\phi},{p}_{\phi} ) = {\cal N} \left( \frac{\di\mathcal{H}}{\di p_{a}}, -\frac{\di\mathcal{H}}{\di a}, \frac{\di\mathcal{H}}{\di p_{\phi}}, -\frac{\di\mathcal{H}}{\di\phi} \right).
\end{equation}
with initial data consistent with the Hamiltonian constraint $\mathcal{H}=0$.  

A more efficient procedure involves identifying a function on phase space as a time variable and solving the Hamiltonian constraint explicitly.  The net effect  is a move from a 4-dimensional system with a Hamiltonian constraint to an unconstrained 2-dimensional system with a non-zero ``physical'' Hamiltonian.  The equations of motion obtained in either approach are of course equivalent (up to time re-parameterizations) give rise to the same classical dynamics.

 \subsection{Time gauges}\label{sec:time gauges}
 
 The reduced action is time reparametization invariant. Therefore physical degrees of freedom are identified after fixing a gauge and solving the Hamiltonian constraint. Canonical  time gauge fixing is the imposition
 \be
  t= f(\text{phase space variables}).
  \ee
   For cosmology it is perhaps natural to consider the class of canonical gauge conditions
 \be\label{eq:time gauge}
 t = f(a). 
 \ee 
(Note that $f$ must carry the units of time; i.e., inverse mass.)  In Dirac's terminology, such a condition must be second class with the Hamiltonian constraint ${\cal H}=0$ (otherwise it does not fix gauge), and so one must either use Dirac brackets, or solve the pair of constraints explicitly. 

 The latter route is straightforward for cosmology.  The gauge fixed reduced action is given by
\begin{equation}
	S^\text{GF}_\text{R} = V_0 \int dt \left[ p_a\dot{a} +p_\phi \dot{\phi} - {\cal N} {\cal H} \right] \bigg|_{\gauge}.	
\end{equation}
Differentiating both sides of (\ref{eq:time gauge}) with respect to $t$ we obtain:
\begin{equation}
	1 = f'(a) \dot{a} \quad \Rightarrow \quad  \dot{a} = 1/f'(a),
\end{equation}
We can use this to write
\begin{align}
	S^\text{GF}_\text{R} &  = V_0 \int dt \left[ \frac{p_a}{f'(a)} +p_\phi \dot{\phi} -\mathcal{N}\mathcal{H} \right] \bigg|_{\gauge} \nn \\
	& = V_0 \int dt \, \left(p_\phi \dot{\phi} - \Hp \right),	
\end{align} 
where we have identified the physical Hamiltonian as
\be
\Hp = -\frac{p_a}{f'(a)} \bigg|_{\gauge}.
\ee
Notice that $\Hp$ is the negative of the momentum conjugate to the time function since 
\be
\left\{ f(a), \frac{p_a}{f'(a)} \right\} = 1. 
\ee
Explicitly solving the Hamiltonian constraint for $p_{a}$ (and restricting ourselves to expanding universes with $p_{a}<0$) allows us to write the physical Hamiltonian as
\be
\Hp = \frac{ 6 \Mp a^2}{f'(a)} \   \sqrt{\frac{\rho_\phi}{3} } = \frac{ 6 \Mp }{f'(a)} \   \sqrt{\frac{ \mathcal{H}_{\phi} a}{3} }=\frac{ 6\Mp^{2}a^2}{f'(a)} H.
\ee
In each of these expressions, it is understood that the scale factor is to be written as an explicit function of $t$ via $a = f^{-1}(t)$ and the Hubble parameter is to be written in terms of $(\phi,p_{\phi},t)$ using the Friedmann equation (\ref{Feqn}).
 
 Lastly, the requirement that the gauge condition is preserved in time leads to the condition fixing the lapse function:
  \be
 \dot{t} = 1 =  \dot{f}(a) = {\cal N}\{f(a), {\cal H}  \} 
 \ee
 Rearranging, we obtain
 \be
 {\cal N} = \frac{1}{  \{ f(a), {\cal H} \}} \bigg|_\gauge  = \frac{\Mp}{f'(a)}\ \sqrt{\frac{3a}{{\cal H}_\phi}  }.
 \ee
 This gives ${\cal N}(\phi,p_\phi,t)$ corresponding to the canonical time gauge $t=f(a)$ on the constraint surface. 
 
As examples, let us consider the following two gauges 
 \begin{equation}\label{eq:gauges}
 	t = \frac{1}{\Mp} \begin{cases} (a/a_0)^{3}, \\ \ln(a/a_{0}), \end{cases}
 \end{equation}
 where $a_{0}$ is a constant that picks a particular (covariantly-defined) reference epoch.  For example, the reference epoch could be taken as the hypersurface where the Hubble parameter takes on a certain value.  Since $a_{0}$ is associated with a physically defined instant of time, we have the following behaviour under spatial dilations (\ref{eq:re-scalings}):
\begin{equation}
	a_{0} \mapsto \kappa^{-1/3} a_{0} \quad \Rightarrow \quad t \mapsto t.
\end{equation}
That is, our choice of time is assumed to be invariant under spatial dilations.
 
 The first of these time gauge in (\ref{eq:gauges}) gives 
 \be
 \Hp =  2\Mp^{2} \sqrt{ \frac{\mathcal{H}_{\phi} a_{0}^{3}}{3\Mp t}} = 2\Mp^{3} a_{0}^{3} H.
 \ee
 The corresponding lapse function is
  \be
  {\cal N} =  \sqrt{ \frac{\Mp^{3}a_{0}^{3}}{3t {\cal H}_\phi} }.
  \ee
 
The second gauge choice is can be interpreted as follows:  Consider two cosmological epochs where the time coordinate and the scale factor are $(t_{1},a_{1})$ and $(t_{2},a_{2})$, respectively.  Then, the number of e-folds of cosmological expansion between the two epochs is
\begin{equation}
	\Delta N = \ln \frac{a_{2}}{a_{1}} = \Mp (t_{2}-t_{1}) = \Mp \Delta t.
\end{equation}
That is, $t$ measures the number of e-folds of expansion along a given trajectory in this gauge, and we hence call it the ``e-fold'' time.  We define an absolute e-fold scale by
\begin{equation}\label{eq:e-fold def}
	N = \Mp t = \ln \frac{a}{a_{0}}
\end{equation}  
so that $N = 0$ corresponds to $a = a_{0}$.\footnote{We caution the reader that the e-fold time $N$ should not be confused with ADM lapse function $\mathcal{N}$.  Also, unlike some of the cosmological literature, we choose $N$ to increase to the future; i.e., $da/dN > 0$.}  In this gauge, we have
 \be
 \Hp = 6\Mp^{2} \sqrt{ \frac{\mathcal{H}_{\phi}a_{0}^{3} e^{3\Mp t} }{3} } = 6 \Mp^{3} a_{0}^{3} e^{3\Mp t} H. \label{efoldH}
 \ee  
 and lapse 
 \be
 {\cal N} =  \Mp^{2} \sqrt{\frac{3a_0^3e^{3\Mp t}}{{\cal H}_\phi}}.
 \ee
 The last two equations may equally well be written in terms of the number of e-folds $N$ using (\ref{eq:e-fold def}); we do this below and use the scalar energy density $\rho_\phi$ instead of the Hamiltonian ${\cal H}_\phi$. 
 
Our goal in the following sections is to study this model in the semiclassical approximation,  which begins with  the effective Hamiltonian constraint 
\be
{\cal H} =  - \frac{p_a^2}{12a\Mp^{2}} +  \left\langle \psi \Bigg|  \frac{\hat{p}_\phi^2}{2a^3} + a^3 V(\hat{\phi}) \Bigg| \psi\right\rangle(\bar{\phi},\bar{p}_\phi), 
\ee
 where $|\psi\rangle$ is a semiclassical state of specified width peaked at a scalar field phase space point $(\bar{\phi},\bar{p}_\phi)$ (which we henceforth write without the bars), and the operators are defined in the polymer quantization prescription. This effective system in the e-fold time gauge $\Mp t=\ln (a/a_0) \equiv N$ defined above is described by the Hamiltonian (\ref{efoldH}).  Defining the effective energy density 
 \be
 \rho_{\text{eff}} = \langle\psi|  \hat{{\rho}}_\phi | \psi \rangle \label{rhoeffdefn}
 \ee
   this Hamiltonian takes the form  
 \be
 \Hp = 6\Mp^{3} a_0^{3} e^{3N} \sqrt{\frac{  {\rho}_{\text{eff}}}{3\Mp^{2}}}. \label{efoldH2}
 \ee
 It remains to calculate $\rho_{\text{eff}}$, to which we now turn.

\section{Polymer quantization of the scalar field}\label{sec:polymer}

Polymer quantization of the scalar field has been studied by several authors both at the formal level and applied to physical systems. There are two versions of it depending on whether momenta or configuration variables are diagonal. Unlike in Schrodinger quantization there is no Fourier transform connecting these, as the basic classical variables are  quite different. The approach we take was introduced in \cite{Husain:2010gb}, which is the one we use in this work.  

Quantization of a scalar field on a background metric 
\be
ds^2 = -{\cal N}^2 dt^2 + (  dx^a + N^a dt   ) (  dx^b + N^b dt ) q_{ab}   
\ee
in this approach starts with the introduction of a pair of non-canonical phase space variables whose algebra resembles the holonomy-flux  algebra used in LQG. These variables are 
\begin{equation}
\Phi_f \equiv \int d^3 x ~ \sqrt{q} f(x) \phi(x) , \quad U_{\lambda} \equiv \exp \left(\frac{i \lambda p_{\phi}}{\sqrt{q}} \right),
\end{equation}
where $f(x)$  is a smearing function. The parameter $\lambda$ is a spacetime constant with dimensions of $ (\text{mass})^{-2} $. These variables satisfy the Poisson algebra
\begin{equation}
\label{eq_pa}
\{ \Phi_f , U_{\lambda} \} = if\lambda U_{\lambda},
\end{equation}

Specializing  to the FRW spacetime with  line element
\begin{equation}\label{eq:FRW}
	ds^{2} = -\mathcal{N}^{2}(t) \, dt^{2} + a^{2}(t) (dx^{2}+dy^{2}+dz^{2}), 
\end{equation}
we can set $f(x)=1$ because of homogeneity, so  these variables become 
\begin{equation}
\Phi = V_0 a^3 \phi , \,\,\,\, U_{\lambda} = \exp \left(\frac{i \lambda p_{\phi}}{a^3} \right),
\end{equation}
Their Poisson bracket is the same as in ($\ref{eq_pa}$).  

Quantization proceeds by realizing the Poisson algebra ($\ref{eq_pa}$) as a commutator algebra on a suitable Hilbert space. The choice for polymer quantization 
has the ``configuration'' basis $ \{ \ket{\mu} ,\  \mu \in \mathds{R}  \} $ with the inner product
\begin{equation}
\braket{\mu^\prime}{\mu} = \delta_{\mu, \mu^{'}},
\end{equation}
where $\delta$ is the generalization of the Kronecker delta to the real numbers. This is the  central difference from the Schrodinger quantization. Explicitly the inner product is 
\be
\langle \mu| \mu'\rangle := \lim_{T\rightarrow \infty} \frac{1}{2T} \int_{-T}^T dx \ e^{-ix\mu} e^{ix \mu'} = \delta_{\mu,\mu'}.
\ee
 Plane waves are normalizable in this inner product. 

The operators $ \hat{\Phi} $ and $ \hat{U}_{\lambda} $ have the action
\begin{equation}
\hat{\Phi}\ket{\mu} = \mu \ket{\mu}, \,\,\,\, \hat{U}_{\lambda} \ket{\mu} = \ket{\mu + \lambda},
\end{equation}
$ \ket{\mu} $ is an eigenstate of the field operator $ \hat{\Phi}$, and $\hat{U}_{\lambda}$ is the generator of field translations. 
As a consequence, the momentum operator does not exist in this quantization because the translation operator is  not weakly continuous in $\lambda$. This may be seen by noting that the limit  
\be
\lim_{\gamma \rightarrow 0}  \frac{1}{\gamma} \langle \mu | U_\gamma - U_0    | \mu\rangle, 
\ee 
which could define the momentum operator from the translation operator $U_\lambda$, does not exist. Nevertheless the momentum operator may be defined indirectly as 
\begin{equation}
\label{P_def}
p_{\phi}^{\lambda} := \frac{a^3}{2i \lambda} ( U_{\lambda} - U_{\lambda}^{\dag} ), 
\end{equation}
a form  motivated by the expansion of  $U_\lambda$.  This   modifies the kinetic energy, and constitutes  the origin of ``polymer corrections,'' while yielding in a suitable limit the standard Schrodinger results \cite{Husain:2010gb,Hossain:2010wy}.

 For polymer corrected cosmological dynamics we wish to calculate the scalar energy density  
\be 
\rho_{\text{eff}} = \frac{1}{a^3} \langle {\cal H}_\phi \rangle 
\ee
  for the  Gaussian coherent state peaked at the phase space values $(\phi , p_{\phi})$. Before doing so we fix the polymer energy scale by setting 
 $\lambda =\lambda_*\equiv M_\star^{-2}$. The state is 
 \begin{align} \nonumber
\ket{\psi} & = \frac{1}{\mathfrak{N}} \sum_{-\infty}^{\infty} c_{k} \ket{\mu_k}, \\ c_{k} & \equiv \exp \left[ - \frac{(\phi_{k} - \phi)^2}{2 \sigma^2} \right] \exp(- i p_{\phi} \phi_{k} V_{0}),
\end{align}
where $\phi_k \equiv \mu_k/V_0 a^3$ is an eigenvalue of the scalar field operator (rather than its integrated version $\hat{\Phi}_k$).  

The effective density $\rho_{\text{eff}}$ was computed for the zero potential case in \cite{Hossain:2010wy}, so we quote the relevant parts. 
The normalization constant is calculated by approximating the sum by an integral to give 
\be
\mathfrak{N} = \sum |{c_{k}}|^2 \simeq V_{0} a^3 \sigma \sqrt{\pi},
\ee
and 
\be
\langle U_{\lambda*} \rangle =  e^{i\Theta} e^{-\Theta^2 /4 \Sigma^2},
\ee
where 
\be
\Theta \equiv \frac{p_\phi}{M^2_*a^3}, \ \ \ \Sigma= V_0\sigma p_\phi
\ee
are variables invariant under the coordinate scale changes $x\rightarrow lx$. Taken together these give
\be
\rho_{\text{eff}} = \frac{M_\star^4}{4}\left(   1- e^{-\Theta^2/\Sigma^2} \cos 2\Theta   \right) + \langle V(\hat{\phi}) \rangle.
\label{rhoeff}
\ee
We study the quadratic potential $V(\phi) = m^2 \phi^2/2$.   Its expectation value in this state is 
 \begin{equation}
\langle V \rangle = \frac{1}{2} m^{2} \left( \phi^{2} + \frac{\sigma^{2}}{2} \right).
\end{equation}
The nonzero quantum width $\sigma$ of the semi-classical state in the $\phi$ direction ensures that $\langle V(\phi) \rangle \ne V(\phi)$. 

Let us examine the energy density  (\ref{rhoeff}) to see if the width correction to the potential plays a significant role. For $\Theta/\Sigma \gg1$, the very small universes regime,  we have $\rho_{\text{eff}} \rightarrow M_\star^4/4$; this gives the early time polymer phase, which is almost exactly deSitter. For $\Theta/\Sigma <1$, we have 
\be
\rho_{\text{eff}}  \sim \frac{ p^2_\phi}{2a^6}\left[ 1-  \frac{1}{(\sigma V_oa^3)^2M_\star^4}  \right] + \frac{1}{ (2\sigma V_oa^3)^2}  + \langle V \rangle. 
\ee
This exhibits the polymer scale ($M_\star$) and width  ($\sigma$) corrections.  (In standard quantum mechanics $M_\star \rightarrow \infty$, and we would have only the width corrections.)  Therefore for sufficiently early times, the kinetic correction to $\rho_{\text{eff}}$ dominates the potential energy correction.   It follows that  we can set $\langle V(\phi) \rangle \approx V(\phi)$.  Finally for $\Theta/\Sigma \ll 1$,  the energy density tends to its classical value with the addition of a width correction that acts as a small cosmological constant.

\section{Semiclassical dynamics}\label{sec:EOMs}

\subsection{Equations of motion}

In this section we study the dynamics that follows from the effective e-fold time physical Hamiltonian density (\ref{efoldH2}).  The equations of motion are given by Hamilton's equations
\begin{equation}
	\Mp \frac{d}{dN} (\phi,{p}_{\phi} ) = \left( \frac{\di\Hp}{\di p_{\phi}}, -\frac{\di\Hp}{\di\phi} \right).
\end{equation}
In terms of the dimensionless quantities defined by 
\begin{align}\nonumber
	 \varphi & := \phi/\Mp, & p_{\varphi} & := p_\phi/ \Mp^{2} a_{0}^{3} , & \gamma  & := M_\text{Pl}/M_{\star}, \\
	 \alpha & := \Mp^{2} \sigma V_{0}, & \mathcal{V} & := \langle V \rangle/\Mp^{4}, &  \tilde{H} & := H/\Mp,
\end{align}
the equations of motion are
\begin{subequations}\label{eq:EOMs}
\begin{align}
	\frac{d\varphi}{d\efolds} & = \frac{1}{2\tilde{H} \gamma^{2}} \exp \left( -\frac{\gamma^{4}}{\alpha^{2} a_{0}^{6} e^{ 6\efolds}} \right) \sin\left( \frac{2 \gamma^{2} p_\varphi}{e^{3\efolds}} \right), \\
	\frac{dp_{\varphi}}{d\efolds} & = -\frac{e^{3\efolds}}{\tilde{H}} \frac{d\mathcal{V}}{d\varphi} .
\end{align}
\end{subequations}
 The dimensionless Hubble parameter is explicitly given by	
\begin{equation}	\label{eq:dimensionless Hubble}
	 \tilde{H} = \sqrt{\frac{\mathcal{V}}{3}  + \frac{1}{12\gamma^{4}} \left[ 1 -  \exp \left( -\frac{\gamma^{4}}{\alpha^{2}a_{0}^{6} e^{ 6\efolds}} \right) \cos\left( \frac{2 \gamma^{2} p_\varphi}{e^{3\efolds}} \right) \right]}.
\end{equation}

We note that under spatial dilations (\ref{eq:re-scalings}), the only quantities appearing in (\ref{eq:EOMs}) or (\ref{eq:dimensionless Hubble}) that are \emph{not} invariant are $\alpha$ and $a_{0}$, which transform as:
\begin{equation}
	(\alpha,a_{0}) \mapsto (\kappa \alpha, \kappa^{-1/3} a_{0}). 
\end{equation}
This of course implies that the equations themselves are invariant since only the combination $\alpha a_{0}^{3}$ appears.  For numeric simulations, we fix the spatial dilation gauge by making the choice $\alpha a_{0}^{3} = \Mp^{2} \sigma V_{0} a_{0}^{3} = 1$.  Physically, this involves fixing the $N=0$ hypersurface as the epoch when the semiclassical width of the smeared field $\Phi = V_{0} a^{3} \phi$ is equal to $\Mp^{-2}$.  Solutions of (\ref{eq:EOMs}) for $\alpha a_{0}^{3} \ne 1$ can be generated from $\alpha a_{0}^{3} = 1$ solutions by the transformations:
\begin{equation}
	N \mapsto N + N_{0}, \quad p_{\varphi} \mapsto e^{3N_{0}} p_{\varphi}.
\end{equation}
 
\subsection{Simulations and phase portraits}\label{sec:simulations}

We begin our study of the solutions of the equations of motion (\ref{eq:EOMs}) by performing numeric simulations.  We limit ourselves to the quadratic potential
\begin{equation}
	V(\phi) = \frac{1}{2} m^{2} \phi^{2}.
\end{equation}
Neglecting semi-classical corrections to the expectation value of $\phi^{2}$, this yields the dimensionless potential
\begin{equation}
	\mathcal{V} = \frac{1}{2} \delta^{2} \varphi^{2}, \quad \delta = \frac{m}{\Mp}.
\end{equation}
Our purpose is to gain a qualitative understanding of the behaviour of solutions, which we will analytically rationalize in subsequent sections.

\begin{figure}
\begin{center}
\includegraphics[width=\columnwidth]{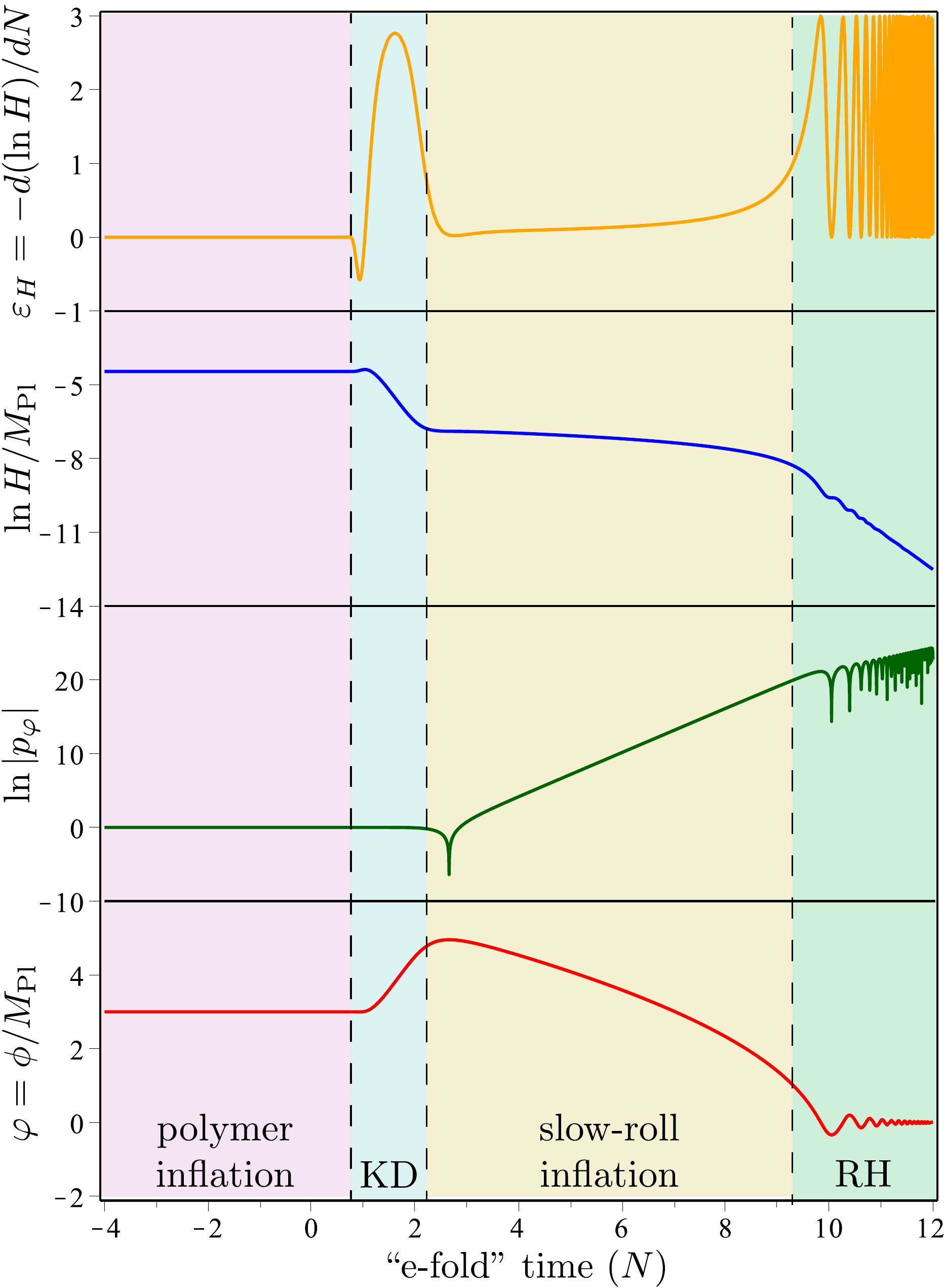}
\end{center}
\caption{Simulation results for the quadratic potential with $(\gamma,\delta,\varphi_{0},p_{0}) = (5,5\times10^{-4},3,1)$.  Here, ``KD'' stands for ``kinetic domination'' and ``RH'' stands for ``re-heating''.}\label{fig:typical}
\end{figure}
In figure \ref{fig:typical}, we plot the output of a single simulation.  Qualitatively, we see four district phases of cosmological evolution:
\begin{enumerate}[(i)]

\item An early time phase where the Hubble factor as well as the scalar amplitude and its conjugate momentum are constant.  The Hubble slow-roll parameter $\varepsilon_{H}$ is very close to 0, implying de Sitter-like inflation.  We call this phase the ``polymer inflation'' epoch.

\item A phase where the Hubble factor appears to decay like $e^{-3N} \propto a^{-3}$, the scalar field depends linearly on $N$, and $\varepsilon_{H}$ approaches 3.  We call this the kinetic domination (``KD'') phase.

\item A ``slow-roll inflation'' phase where the Hubble factor varies slowly, with $\varepsilon_{H}<1$ but not exactly 0; implying quasi-de Sitter expansion.

\item A post-inflation reheating (``RH'') phase where $\varphi$ oscillates about $0$.
	
\end{enumerate}
The last two phases are familiar from ordinary quadratic inflation, while the first is truly distinct from the standard treatment.  

In order to gain a sense of which features shown in figure \ref{fig:typical} are generic, we conducted numerous other simulations with different choices of parameters.  These allow us to empirically conclude that the polymer inflation and reheating phases are always present in solutions of (\ref{eq:EOMs}), however the existence and length of the slow-roll and kinetic-dominated phases depend on the specific choices made.\footnote{As discussed in detail in Sec.\ \ref{sec:early} below, the polymer inflation phase is a robust feature of \emph{all} solutions as long as simulations are started early enough; i.e., when $\gamma^{2} \gg e^{3N}$.  This is a simple consequence of the $N \rightarrow -\infty$ limit of (\ref{eq:EOMs}), and holds irrespective of parameter and initial data choices.}

\begin{figure}
\begin{center}
\includegraphics[width=\columnwidth]{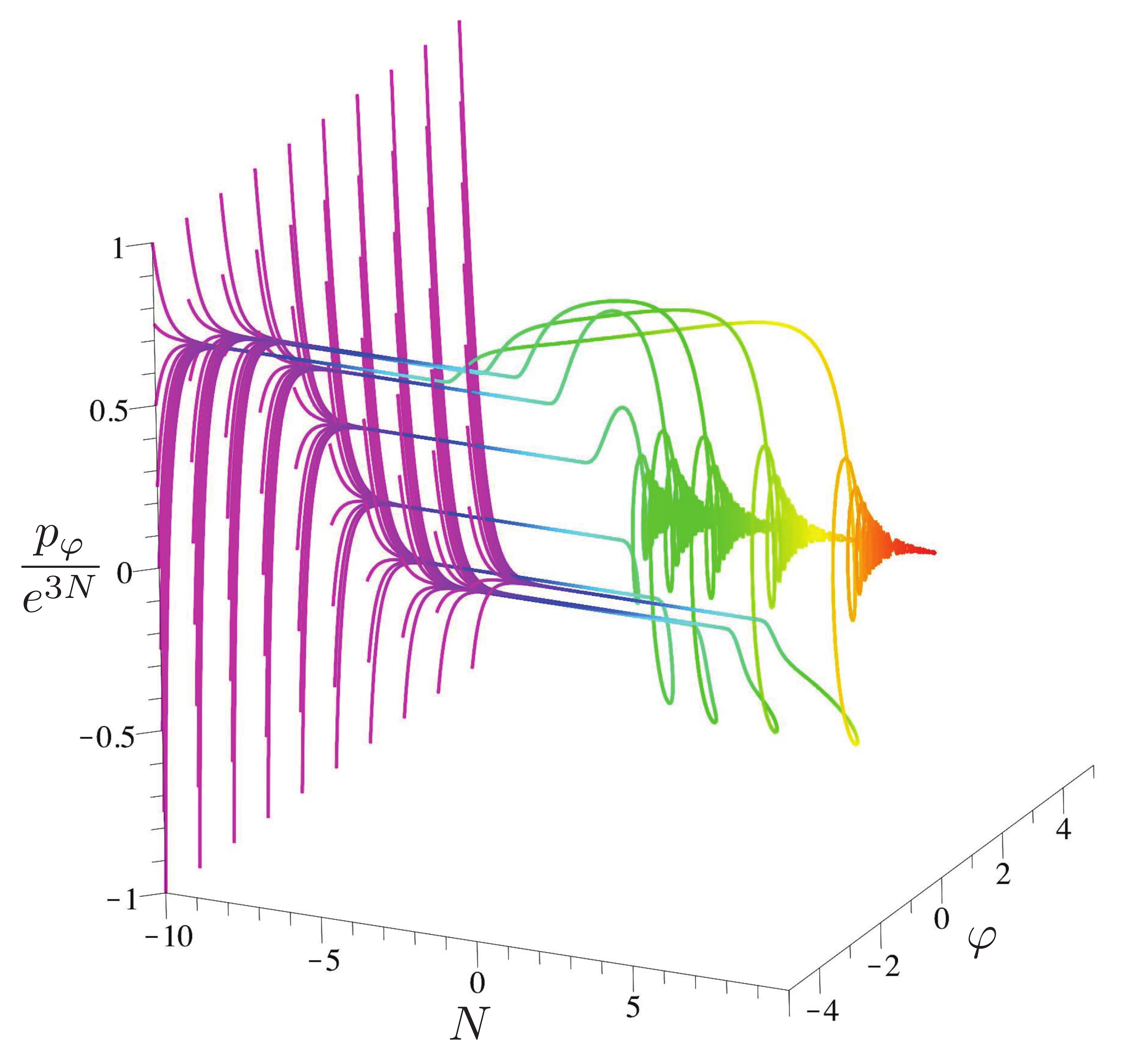}
\end{center}
\caption{3-dimensional phase portrait of the system (\ref{eq:EOMs}) assuming a quadratic potential with $(\gamma,\delta) = (0.5,1.0)$.}\label{fig:phase3D}
\end{figure}
\begin{figure*}
\begin{center}
\includegraphics[width=\textwidth]{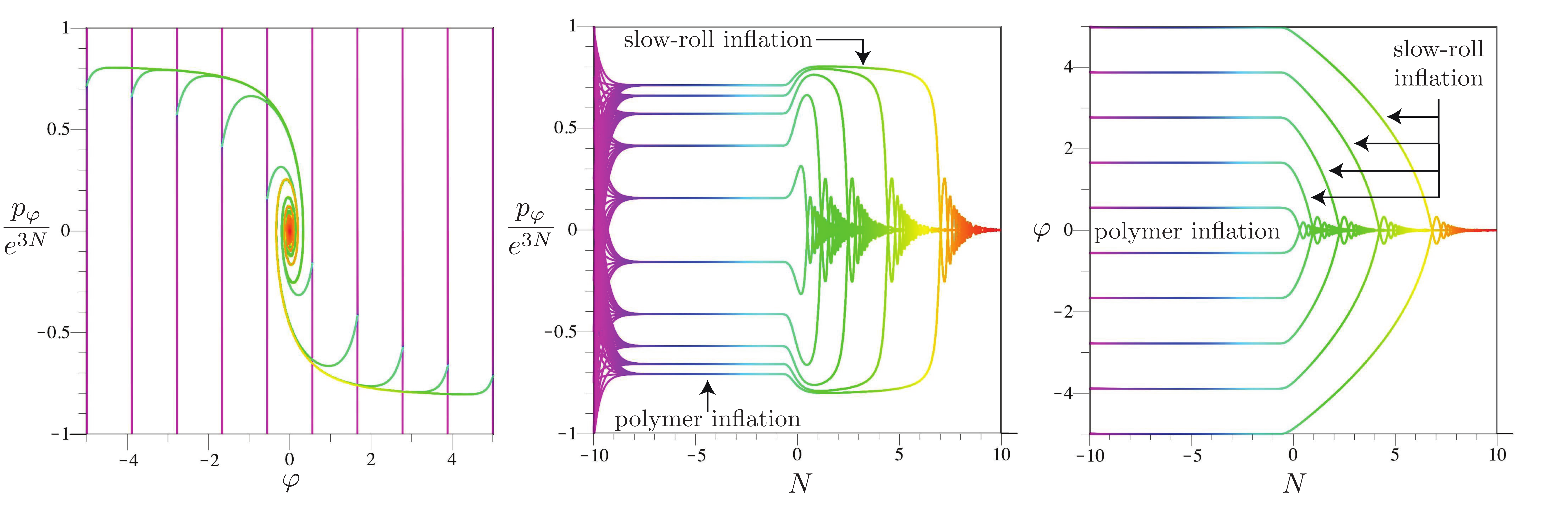}
\end{center}
\caption{Projection of the 3-dimensional phase portrait in figure \ref{fig:phase3D} onto coordinate planes}\label{fig:phase2D}
\end{figure*}
To further illustrate qualitative features of solutions, we plot cosmological trajectories through the 3-dimensional ($\efolds$,$\varphi$,$p_{\varphi}$) phase space in figures \ref{fig:phase3D} and \ref{fig:phase2D}.  For ease in visualization, we plot $p_{\varphi}/e^{3\efolds}$ instead of $p_{\varphi}$.  Initial data for the trajectories is sampled on a rectangular grid in the $(\varphi,p_{\varphi}/e^{3\efolds})$ plane when $\efolds = -10$.  We see that soon after the start of the simulation, all trajectories rapidly approach an attractor manifold (which we will discuss further in \ref{sec:early}).  They remain close to this surface until the end of polymer inflation when $\gamma^{2}/e^{3\efolds} \sim 1$.  This behaviour is also clearly exhibited in the middle panel of figure \ref{fig:phase2D}---which shows the projection of the trajectories in figure \ref{fig:phase3D} onto each of the phase space coordinate planes---where we see the trajectories converge exponentially to the attractor at early times.  Also apparent in this plot is that none of the simulations demonstrate a clear kinetic-dominated phase, and the duration of the slow-roll inflation phase depends on initial conditions, particularly the choice of $\varphi_{0}$.  We will comment on these observations below.

In the following subsections, we derive analytic formulae describing the polymer, slow-roll, and kinetic dominated phases of evolution apparent in figures \ref{fig:typical}--\ref{fig:phase2D}.
 
\subsection{The polymer inflation phase}\label{sec:early}

The polymer inflation phase evident in simulations represents the early time limit of the systems dynamics.  In this limit we assume
\begin{equation}
	\frac{\gamma^{2}}{e^{3\efolds}} \gg 1.
\end{equation}
Now, the equations of motion (\ref{eq:EOMs}) reduce to
\begin{equation}
	\frac{d\varphi}{d\efolds} = 0, \quad \frac{dp_{\varphi}}{d\efolds} = -\frac{e^{3\efolds} \mathcal{V}'(\varphi)}{\tilde{H}}.
\end{equation}
with
\begin{equation}	
	 \tilde{H}^{2} = \frac{1}{3} \mathcal{V}(\varphi)  + \frac{1}{12\gamma^{4}}.
\end{equation}
The solution is
\begin{equation}\label{eq:early time solutions}
	\varphi = \varphi_{0}, \quad \frac{p_{\varphi}}{e^{3N}} = \frac{p_{0}}{e^{3N}} - \frac{2\gamma^{2}\mathcal{V}'(\varphi_{0})}{\sqrt{12\gamma^{4}\mathcal{V}(\varphi_{0}) + 3}}.
\end{equation}
with $\varphi_{0}$ and $p_{0}$ representing constants of integration.  Since $\varphi$ is constant in this limit, the Hubble factor is also constant,  implying the universe undergoes de Sitter-like expansion.

The appearance of the early time attractor manifold seen in figures \ref{fig:phase3D} and \ref{fig:phase2D} is now easy to rationalize from the solutions (\ref{eq:early time solutions}):  We see that irrespective of our choice of initial data, the solution of the equation of motion approaches the surface
\begin{equation}
	\frac{p_{\varphi}}{e^{3N}} + \frac{2\gamma^{2}\mathcal{V}'(\varphi)}{\sqrt{12\gamma^{4}\mathcal{V}(\varphi) + 3}},
\end{equation}
exponentially quickly during the polymer inflation phase. 
 
\subsection{Slow-roll phase}\label{sec:slow roll}

The slow-roll phase of evolution will take place after the end of the polymer inflation phase, which means we can take
\begin{equation}
	\frac{\gamma^{2}}{e^{3\efolds}} \ll 1.
\end{equation}
Hence, the equations of motion (\ref{eq:EOMs}) reduce to
\begin{subequations}
\begin{align}
	\frac{d\varphi}{d\efolds} & = \frac{1}{2\tilde{H} \gamma^{2}} \sin\left( \frac{2 \gamma^{2} p_\varphi}{e^{3\efolds}} \right), \\
	\frac{dp_{\varphi}}{d\efolds} & =  -\frac{e^{3\efolds}}{\tilde{H}} \mathcal{V}'(\varphi),
\end{align}
\end{subequations}
with
\begin{equation}
	 \tilde{H}^{2} = \frac{\mathcal{V}}{3}  + \frac{1}{6\gamma^{4}} \sin^{2}\left( \frac{\gamma^{2} p_\varphi}{e^{3\efolds}} \right).
\end{equation}
To obtain the Hamilton-Jacobi form of the equations of motion, we note that the simulations of section \ref{sec:simulations} suggest that $\varphi$ is a monotonic function during slow-roll, so we can use it as an effective time coordinate.  Then, it is fairly easy to show
\begin{subequations}
\begin{gather}\label{eq:HJ 1}
	\frac{d\tilde{H}}{d\varphi} = -\frac{1}{4\gamma^{2}} \arcsin \left( 2\gamma^{2} \tilde{H} \frac{d\varphi}{dN} \right), \\
	\tilde{H}^{2} - \frac{1}{3} \mathcal{V} = \frac{1}{6\gamma^{4}} \sin^{2}\left( 2\gamma^{2} \frac{d\tilde{H}}{d\varphi} \right). \label{eq:HJ 2}
\end{gather}
\end{subequations}
Expanding the right had side of each of these to leading order in $\gamma$ leads to the standard Hamilton-Jacobi equations of single field inflation. (See e.g.\cite{Spalinski:2007kt} for a similar expansion from brane inflation.)

The slow-roll paradigm is that the Hubble parameter is approximately constant during the inflating period; i.e., $|d\tilde{H}/d\varphi| \ll 1$.  Using this assumption, we can solve (\ref{eq:HJ 2}) iteratively for $\tilde{H} = \tilde{H}(\varphi)$.  Restoring units, we obtain
\begin{equation}\label{eq:slow roll Friedmann}
	H^{2} = \frac{V}{3\Mp^{2}} \left[ 1 + \frac{\varepsilon_{V}}{3}  + \frac{2\varepsilon_{V} \eta_{V}}{9}   - \frac{\varepsilon_{V}^{2}}{3} \left(1 + \frac{2V}{9\M^{4}}  \right)  + \cdots   \right],
\end{equation}
where we have defined the standard potential slow-roll parameters
\begin{subequations}
\begin{align}
	\varepsilon_{V} & =  \frac{1}{2\mathcal{V}^{2}} \left( \frac{d\mathcal{V}}{d\varphi} \right)^{2}=  \frac{\Mp^{2}}{2V^{2}} \left( \frac{dV}{d\phi} \right)^{2},\\
	\eta_{V} & =  \frac{1}{\mathcal{V}} \frac{d^{2}\mathcal{V}}{d\varphi^{2}} =  \frac{\Mp^{2}}{V} \frac{d^{2}V}{d\phi^{2}} .
\end{align}
\end{subequations}
In equation (\ref{eq:slow roll Friedmann}) and below, ``$\cdots$'' indicates terms higher order in slow-roll parameters.  The expansion of the Friedmann equation (\ref{eq:slow roll Friedmann}) suggests that polymer effects come it at second order in slow-roll parameters.  However, we should caution that for this expansion and the ones below to be valid, we require
\begin{equation}\label{eq:consistency}
	\frac{V}{\M^{4}} \lesssim 1.
\end{equation}

Of course the potential slow-roll parameters are not the only ones we can define.  Re-introducing the dimensionless proper time $T$ via
\begin{equation}
	\frac{dN}{dT} = \tilde{H},
\end{equation}
we can define a Hubble slow roll parameter by
\begin{equation}
	\varepsilon_{H} = - \frac{1}{\tilde{H}^{2}} \frac{d\tilde{H}}{dT} = - \frac{d}{dN} \ln \tilde{H}.
\end{equation}
This parameter is useful because of the identity
\begin{equation}
	\frac{1}{a} \frac{d^{2}a}{dT^{2}} = \tilde{H}^{2} ( 1 - \varepsilon_{H}),
\end{equation}
from which we see that inflation only occurs when $\varepsilon_{H} < 1$ and purely deSitter inflation has $\varepsilon_{H}=0$.  Using our previous results, we can write
\begin{equation}
	\varepsilon_{H} = \frac{1}{2\gamma^{2}\tilde{H}^{2}} \frac{d\tilde{H}}{d\varphi} \sin \left( 4\gamma^{2} \frac{d\tilde{H}}{d\varphi} \right) 
\end{equation}
Making use of (\ref{eq:slow roll Friedmann}), this becomes
\begin{equation}
	\varepsilon_{H} = \varepsilon_{V} \left[ 1 - \frac{4}{3} \left(1 + \frac{1}{3} \frac{V}{\M^{4}} \right) \varepsilon_{V} + \frac{2}{3} \eta_{V} + \cdots \right]
\end{equation}

If we solve (\ref{eq:HJ 1}) for $dN/d\varphi$ we can obtain a formula for the number of e-folds of expansion as the scalar field rolls from $\varphi_{1}$ to $\varphi_{2}$.  We obtain
\begin{equation}
	\Delta N = -2\gamma^{2} \int_{\varphi_{1}}^{\varphi_{2}} \frac{\tilde{H}(\varphi)}{\sin [4\gamma^{2} \tilde{H}'(\varphi)]} d\varphi.
\end{equation}
Substituting in (\ref{eq:slow roll Friedmann}), we obtain
\begin{equation}
	\Delta N = \int\limits_{\varphi_{2}}^{\varphi_{1}} \frac{\left[ 1 + \frac{2}{3} \left(1+ \frac{V}{3\M^{4}} \right)\varepsilon_{V} - \frac{1}{3} \eta_{V} + \cdots \right] }{\sqrt{2\varepsilon_{V}}} d\varphi.
\end{equation}

To conclude this subsection, we specialize to the quadratic potential:
\begin{equation}
	V = \frac{1}{2} m^{2} \phi^{2}, \quad \varepsilon_{V} = \eta_{V} = \frac{2\Mp^{2}}{\phi^{2}}.
\end{equation}
The consistency condition (\ref{eq:consistency}) reduces to
\begin{equation}
	\frac{\Mp^{2}m^{2}}{\M^{4}} \lesssim \varepsilon_{V}.
\end{equation}
We find the following:
\begin{subequations}
\begin{align}
	H^{2} & \approx \frac{m^{2}\phi^{2}}{6\Mp^{2}} +\frac{m^{2}}{9}\left( 1 - \frac{2}{9} \frac{m^{2}\Mp^{2}}{\M^{4}} \right),\\
	\varepsilon_{H} & \approx \frac{2\Mp^{2}}{\phi^{2}} \left( 1 - \frac{4}{9} \frac{m^{2}\Mp^{2}}{\M^{4}} \right) - \frac{8\Mp^{4}}{3\phi^{4}}, \label{eq:quad approx 2} \\
	\Delta N & \approx  \left(1 + \frac{4}{9} \frac{m^{2}\Mp^{2}}{\M^{4}} \right) \left( \frac{\phi_{1}^{2} - \phi_{2}^{2}}{4\Mp^{2}} \right) + \frac{1}{3} \ln \frac{\phi_{1}}{\phi_{2}}.\label{eq:quad approx 3} 
\end{align}
\end{subequations}

Assuming that slow roll inflation starts at some $\phi = \phi_{1}$, it will end when $\varepsilon_{H} = 1$.  Neglecting the $\phi^{-4}$ term in (\ref{eq:quad approx 2}), this implies the end of inflation is when
\begin{equation}
	\phi^{2} = 2\Mp^{2} \left( 1 - \frac{4}{9} \frac{m^{2}\Mp^{2}}{\M^{4}} \right).
\end{equation}
Neglecting the logarithmic term in (\ref{eq:quad approx 3}) and working to leading order in $\M^{-4}$ this yields the total number of e-folds of slow roll inflation to be
\begin{equation}\label{eq:Ntot 1}
	\Ntot = \frac{\phi_{1}^{2}}{4\Mp^{2}} \left( 1 + \frac{4}{9} \frac{m^{2}\Mp^{2}}{\M^{4}} \right) - \frac{1}{2}.
\end{equation} 
Now, we would like to relate the field value $\phi_{1}=\Mp\varphi_{1}$ at the onset of slow-roll to the field value during the polymer inflation phase $\varphi_{0}$.  To do so, we must look at the kinetic-dominated phase in between polymer inflation and slow-roll in greater detail.
 
\subsection{Kinetic-dominated phase}\label{sec:kinetic}
 
Since it is conserved at early times, the dimensionless Hubble factor at the end of the polymer inflation phase will be
\begin{equation}	
	 \tilde{H} = \sqrt{\frac{1}{3} \mathcal{V}(\varphi_{0})  + \frac{1}{12\gamma^{4}}}.
\end{equation} 
If we have
\begin{equation}
	\frac{1}{3} \mathcal{V}(\varphi_{0})  \gg \frac{1}{12\gamma^{4}},
\end{equation}
the slow-roll conditions will be satisfied at the end of polymer inflation and the universe will immediately enter into slow-roll.  This is the situation depicted in figures \ref{fig:phase3D} and \ref{fig:phase2D}.  In this scenario, we can take the field value at the start of slow-roll to simply be
\begin{equation}
	\phi_{1} = \Mp \varphi_{0}.
\end{equation}

\begin{figure}
\begin{center}
\includegraphics[width=0.9\columnwidth]{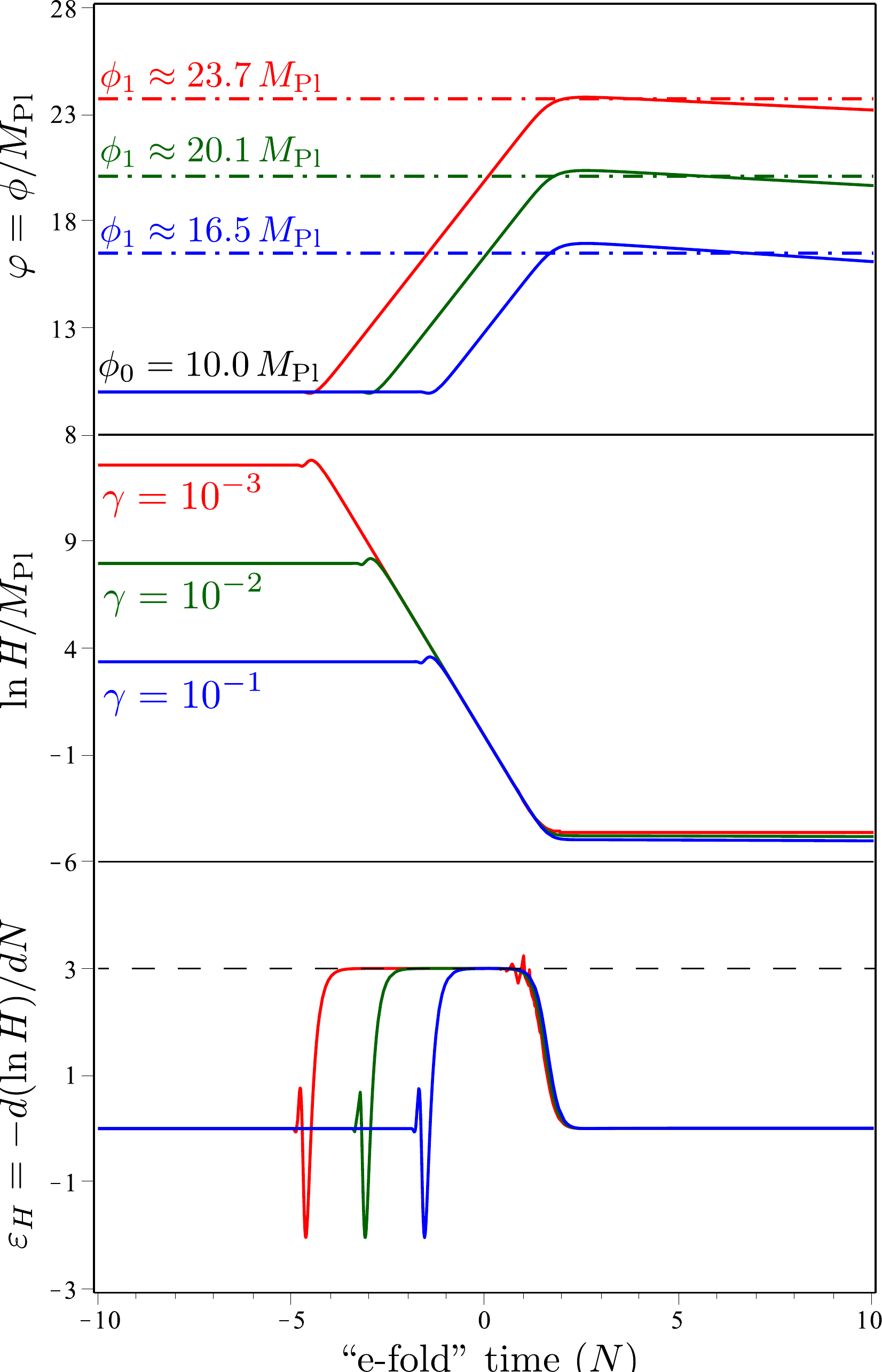}
\end{center}
\caption{Simulation results depicting the kinetic dominated phase of cosmological evolution assuming a quadratic potential.  We take $(\delta,\varphi_{0},p_{0})=(10^{-3},10.0,2.0)$.  During this phase $\varepsilon_{H} \approx 3$ as seen in the lower panel, and the scalar amplitude varies linearly with $N$.  The dash-dot horizontal lines in the top panel show the estimate for the scalar field amplitude at the onset of slow roll inflation based on the numerical solution of equation (\ref{eq:transcendental}) for $\phi_{1}$.  Note how the magnitude of the field excursion $|\phi_{1}-\phi_{0}|$ increases with decreasing $\gamma$.}\label{fig:kinetic}
\end{figure}

However, if we instead have
\begin{equation}\label{eq:KD condition}
	\frac{1}{3} \mathcal{V}(\varphi_{0})  \ll \frac{1}{12\gamma^{4}},
\end{equation}
there must exist a transitional period between polymer and slow-roll inflation.  This is the scenario shown in figure \ref{fig:typical}.  The condition (\ref{eq:KD condition}) implies that potential energy in the the Hubble-factor is sub-dominant, so we can call this the kinetic-dominated phase of evolution.  In figure \ref{fig:kinetic} we show several simulations (with the quadratic potential) depicting the kinetic dominated transition between the polymer and slow-roll inflation epochs.  The main features of the kinetic dominated phase is that the Hubble slow roll parameter $\varepsilon_{H}$ is approximately equal to $3$ implying that the Hubble factor decays like $e^{-3N}$ or $a^{-3}$, and that the scalar field varies linearly with $N$.  This last feature means that the scalar field amplitude at the end of polymer inflation $\phi_{0}$ is not necessarily the same as the amplitude at the start of slow-roll inflation $\phi_{1}$.   Our goal in this subsection is to estimate the functional dependence of $\phi_{1}$ on $\phi_{0}$ and other parameters.

\subsubsection{Case 1: $|p_0| \gtrsim 1$}

Just as in the slow-roll phase above, any kinetic-dominated phase occurs after polymer inflation, and we take
\begin{equation}
	\frac{\gamma^{2}}{e^{3\efolds}} \ll 1.
\end{equation}
Now, if we assume that $|p_{0}| \ge \mathcal{O}(1)$, we may neglect the scalar field potential, leading to the equations of motion:
\begin{equation}\label{eq:kinetic EOMs}
	\frac{d\varphi}{d\efolds} = \frac{1}{2\tilde{H} \gamma^{2}} \sin\left( \frac{2 \gamma^{2} p_\varphi}{e^{3\efolds}} \right), \quad \frac{dp_{\varphi}}{d\efolds} =  0,
\end{equation}
the Friedmann equation
\begin{equation}\label{eq:kinetic Friedmann}
	 \tilde{H}^{2} = \frac{1}{6\gamma^{4}} \sin^{2}\left( \frac{\gamma^{2} p_\varphi}{e^{3\efolds}} \right),
\end{equation}
and Hamilton-Jacobi equations
\begin{subequations}
\begin{gather}\label{eq:KHJ 1}
	\frac{d\tilde{H}}{d\varphi} = -\frac{1}{4\gamma^{2}} \arcsin \left( 2\gamma^{2} \tilde{H} \frac{d\varphi}{dN} \right), \\
	\tilde{H}^{2} = \frac{1}{6\gamma^{4}} \sin^{2}\left( 2\gamma^{2} \frac{d\tilde{H}}{d\varphi} \right). \label{eq:KHJ 2}
\end{gather}
\end{subequations}
Equations (\ref{eq:kinetic EOMs}) imply that $p_{\varphi}$ is constant and that the sign of $d\varphi/dN$ is the same as the sign of $p_{\varphi}$.  This in turn implies that the sign of $d\tilde{H}/d\varphi$ is the opposite of $p_{\varphi}$ via (\ref{eq:KHJ 1}).  This allows us to rewrite (\ref{eq:KHJ 2}) as
\begin{equation}\label{eq:dphi dH}
	\frac{d\varphi}{d\tilde{H}} = -\frac{2\,\text{sgn}(p_{\varphi}) \gamma^{2}}{\arcsin(\sqrt{6} \gamma^{2} \tilde{H})}.
\end{equation}

Now, the kinetic-dominated phase starts when 
\begin{equation}
	\tilde{H} = \tilde{H}_{0} = \sqrt{1/12\gamma^{4}}, \quad \varphi = \varphi_{0}.
\end{equation}
We can take the ending of the the phase to roughly be when the slow-roll condition becomes valid: 
\begin{equation}
	\tilde{H} = \tilde{H}_{1} = \sqrt{\mathcal{V}(\varphi_{1})/3}, \quad \varphi = \varphi_{1}.
\end{equation}
We can hence integrate (\ref{eq:dphi dH}) from $\tilde{H}_{0}$ to $\tilde{H}_{1}$ to obtain the change in $\varphi$ during kinetic domination.  After some straightforward manipulations we obtain:
\begin{equation}
	\varphi_{1} - \varphi_{0} = \frac{\text{sgn}(p_{\varphi})}{\sqrt{3}} \int^{1}_{2\gamma^{2}\mathcal{V}^{1/2}(\varphi_{1})} \frac{dx}{\arcsin(x/\sqrt{2})}.
\end{equation}
If $\tilde{H}_{0} \gg \tilde{H}_{1}$, the lower limit of integration is $\ll 1$.  Under this assumption, the integral is dominated by the portion of the integrand corresponding to $x \ll 1$, and we can use the small argument approximation of $\arcsin$ to obtain:
\begin{equation}
	\varphi_{1} - \varphi_{0} = -\sqrt{\frac{2}{3}} \text{sgn}(p_{\varphi}) \ln[ 2\gamma^{2}\mathcal{V}^{1/2}(\varphi_{1})],
\end{equation}
or written in terms of dimensionful quantities:
\begin{equation}\label{eq:transcendental}
	\phi_{1} = \phi_{0} -\frac{\Mp}{\sqrt{6}} \text{sgn}(p_{\varphi}) \ln\left[ \frac{4V(\phi_{1})}{\M^{4}}\right],
\end{equation}
Unfortunately, for most potentials this will be a highly nonlinear equation to solve for $\varphi_{1}$ that is impossible to solve analytically.

However, for the quadratic potential $V=\frac{1}{2} m^{2} \phi^{2}$ it is fairly easy to obtain a numeric solution.\footnote{Technically speaking, on can write down analytic formula for $\phi_{1}$ in this case in terms of Lambert-W functions, but these have multiple branches which make the expressions difficult to work with.}  Such numerical solutions are compared to simulation results in figure \ref{fig:kinetic}.   Furthermore, one can find a (rather rough) fitting formula to the numerical results.
\begin{equation}\label{eq:Ntot 2}
	\phi_{1} \sim \phi_{0} - \Mp \, \text{sgn}(p_{\varphi}) \left[ 0.8\,\ln \left( \frac{m\Mp}{\M^{2}} \right) + 3.8 \right].
\end{equation}
Interestingly we see that if $m\Mp \ll \M^{2}$, we will have $|\phi_{1} - \phi_{0}| \gg \Mp$.  That is, the kinetic dominated phase can produce very large field values at the onset of slow-roll inflation; this effect is also evident in figure \ref{fig:kinetic}.  

Even if we do not assume $m\Mp \ll \M^{2}$, it is still possible to achieve $|\phi_{1}| \gtrsim \Mp$ if the initial value $\phi_{0}$ is small via the constant 3.8 term in (\ref{eq:Ntot 2}).  From equation (\ref{eq:Ntot 1}), we see that this implies that sub-Planckian initial data can lead to an appreciable amount of slow-roll inflation.  (However, this assumes $|p_{0}| \gtrsim 1$; we discuss small $p_{0}$ is Sec.\ \ref{sec:small p} below.)

Finally, we can also explain the observation from figure \ref{fig:kinetic} that $\varphi \propto N$ and $H \propto e^{-3N}$ during kinetic domination.  Recalling that $p_{\varphi}$ is constant and assuming that $\gamma^{2}p_{\varphi}$ is not extremely large, we are guaranteed that a few e-folds after the start of kinetic domination we have $\gamma^{2}p_{\varphi}/e^{3N} \ll 1$.  Then, it is fairly easy to show that (\ref{eq:kinetic EOMs}) and (\ref{eq:kinetic Friedmann}) yield
\begin{equation}
	\varphi = \sqrt{6} \,\text{sgn}(p_{\varphi})N + \text{constant}, \quad \tilde{H}= \frac{1}{\sqrt{6}} \frac{|p_{\varphi}|}{e^{3N}},
\end{equation}
which are the behaviours seen in Fig.\ \ref{fig:kinetic}.

\subsubsection{Case 2: $|p_{0}| \ll 1$}\label{sec:small p}

\begin{figure}
\begin{center}
\includegraphics[width=\columnwidth]{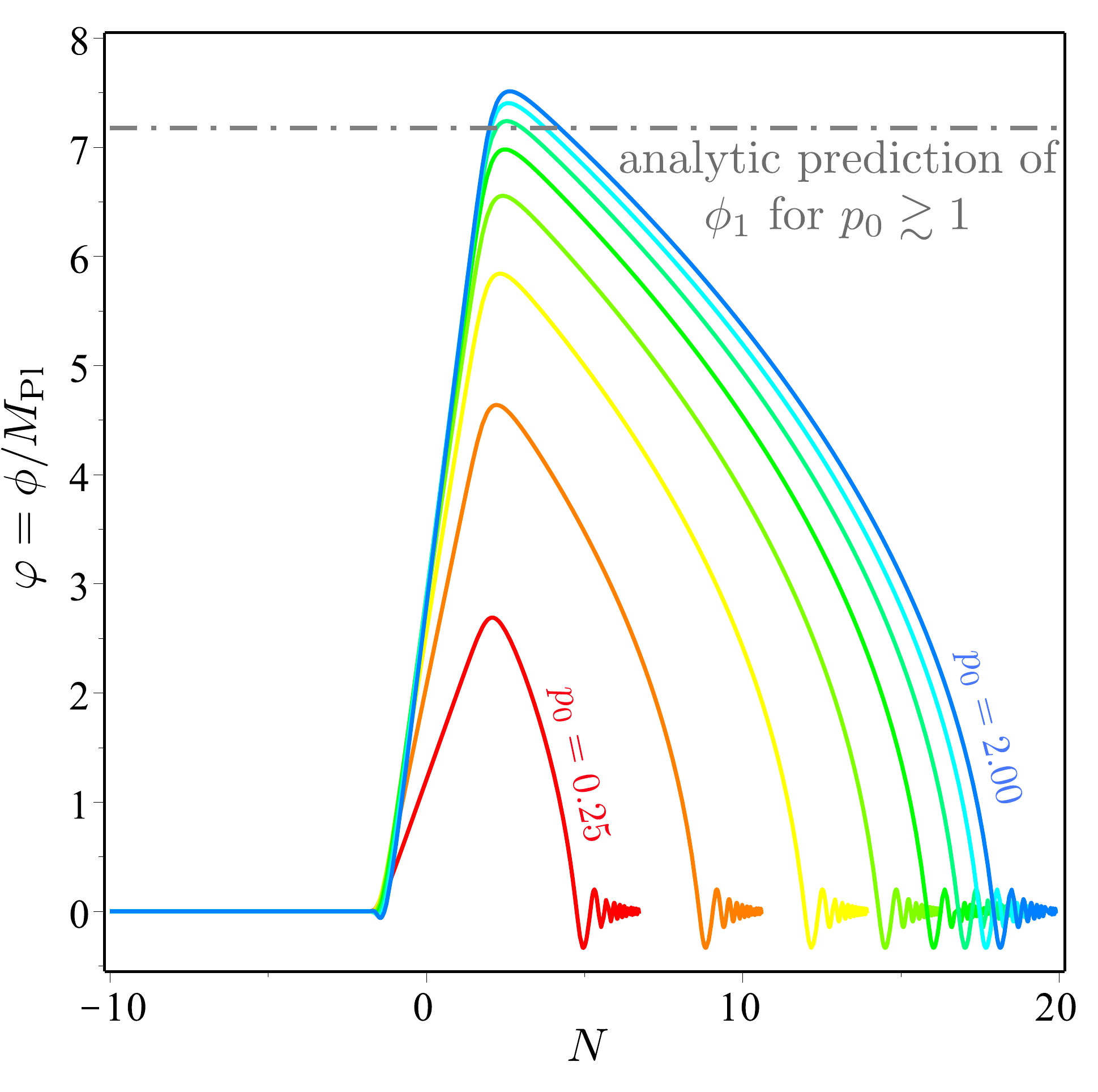}
\end{center}
\caption{Simulations showing the evolution of the scalar field amplitude for sub-Planckian initial data assuming a quadratic potential.  We take $\gamma = 0.1$, $\delta = 10^{-3}$ and an initial scalar field amplitude of $\phi_{0} = 10^{-3} \, \Mp$.  Each curve represents different choices for the initial momentum $p_{0}$, starting with $p_{0} =0.25$ and increasing to $p_{0}=2.00$ in increments of $0.25$.  The dash-dot line is the analytic prediction (\ref{eq:Ntot 2}) for the value of $\phi$ at the end of kinetic domination.  As expected, the maxima in the numeric curves are consistent with the approximation for $p_{0} \gtrsim 1$.}\label{fig:kinetic_p_vary}
\end{figure}
Equation (\ref{eq:Ntot 2}) seems to imply that is possible to have scenarios where the scalar amplitude is sub-Planckian in the polymer inflation phase $|\phi_{0}| \ll \Mp$ yet super-Planckian at the onset of slow-roll inflation $|\phi_{1}| \gtrsim \Mp$.  This is interesting from the point of view of the standard initial condition problem of ordinary inflation: For a quadratic potential, one requires (presumably unnatural) super-Planckian initial data to ensure a sufficient amount of inflation to explain CMB (and other) observations.

However, some caution is required since (\ref{eq:Ntot 2}) was derived under the assumption that $|p_{0}| \gtrsim 1$.  If this condition is not satisfied, then the Hamilton-Jacobi equations (\ref{eq:KHJ 1}) used in the derivation of (\ref{eq:Ntot 2}) are not valid.  For such cases we can resort to numeric simulations like the ones shown in Figure \ref{fig:kinetic_p_vary}.  There, we assume a very small initial scalar amplitude of $\phi_{0} = 10^{-3} \Mp$ and various values of $p_{0}$.  We see that for $|p_{0}| \lesssim 1$, the value of $\phi$ at the onset of slow-roll is not as large as when $|p_{0}| \gtrsim 1$, but it is still super-Planckian.  We also see that the maximum amplitude attained is consistent with the analytic prediction for $|p_{0}| \gtrsim 1$.

To summarize, in the polymer inflation model it is possible for sub-Planckian initial data to lead to significant slow-roll inflation due to the existence of a kinetic dominated phase prior to slow-roll.  We confirmed this via an analytic approximation for large initial values of the scalar momentum and numeric simulations for small initial momentum.


\section{Probability of inflation}\label{sec:e-folds}

A question that has drawn a fair amount of attention in the literature is whether inflation is generic or a result of finely tuned initial conditions.  This problem is sometimes posed by reducing the cosmological dynamics to an effective two dimensional $(\phi,\dot\phi)$ dynamical system \cite{Kofman:2002cj,Remmen:2013eja,Remmen:2014mia}, while other authors have approached it using a Hamiltonian framework without time gauge fixing \cite{Gibbons:1986xk,Gibbons:2006pa,Ashtekar:2009mm,Ashtekar:2011rm,Corichi:2010zp,Corichi:2013kua}.  Here, we will briefly review the non-gauge fixed Hamiltonian formulation, and then propose a new strategy based on the physical Hamiltonian introduced in Section \ref{sec:time gauges}.  

As discussed in Section \ref{sec:Hamiltonian cosmology}, the phase space of single field inflationary cosmology is \emph{a priori} 4-dimensional, and is covered by coordinates $(a,p_{a},\phi,p_{\phi}) \in \mathbb{R}^{4}$.  The Hamiltonian constraint $\mathcal{H}=0$ restricts the cosmological dynamics to an embedded 3-surface in this phase space.  

To assign a probability to inflation, many authors  choose a 2-surface $\mathcal{S}$ within the constraint surface that all phase space trajectories \emph{cross only once}.  This implies a unique mapping of points in $\mathcal{S}$ onto cosmological trajectories.  Given an integration measure $d\sigma$ and a distribution function $\rho$ of trajectories crossing $\mathcal{S}$, this allows one to calculate the probability of inflation:
\begin{equation}\label{eq:P formula}
	P_\text{inf} = {\displaystyle \int_{\mathcal{S}_\text{inf}} \rho \, d\sigma} \Bigg/ {\displaystyle \int_{\mathcal{S}} \rho \, d\sigma},
\end{equation}
where $\mathcal{S}_\text{inf}$ is the subset of $\mathcal{S}$ that maps onto trajectories with more than some specified threshold number of inflationary e-folds.  

Another statistic that is sometimes considered is the expected number of e-folds of inflation \cite{Remmen:2014mia}, given by
\begin{equation}\label{eq:Ntot formula}
	\langle \Ntot \rangle = {\displaystyle \int_{\mathcal{S}} \rho \Ntot \, d\sigma} \Bigg/ {\displaystyle \int_{\mathcal{S}} \rho \, d\sigma},
\end{equation}
where $\Ntot$ is a function that maps points in $\mathcal{S}$ onto the number of inflationary e-folds for the associated cosmological trajectory. 

In the above formulae for calculating $P_\text{inf}$ or $\langle \Ntot \rangle$, several choices must be made: 
\begin{enumerate}

\item One needs to choose the integration measure $d\sigma$.  Much of the literature follows Gibbons, Hawking and Stewart \cite{Gibbons:1986xk} and makes use of the fact that Louiville's theorem picks out a preferred (symplectic) form $\omega = da \wedge dp_{a} + d\phi \wedge dp_{\phi}$ on the full 4-dimensional phase space, which in turn defines a phase space volume element conserved under the Hamiltonian flow.  The measure $d\sigma$ is then defined by the pullback of the symplectic form onto $\mathcal{S}$.  

\item Ideally, the probability distribution $\rho$ should be provided to us by quantum gravity or some other fundamental theory, but in the absence of such a theory one is forced to make \emph{ad hoc} assumptions.  The simplest of these comes from Laplace's ``principle of indifference'', which essentially states that without any information to the contrary, one should choose $\rho$ to have the least structure possible, which suggest the choice $\rho = 1$ used by several authors.  (We remark that the choice of $\rho$ is not disconnected with the choice of measure $d\sigma$ since only the combination $\rho\,d\sigma$ appears in probabilities or expectation values.)

\item Finally, there is the choice of the surface $\mathcal{S}$.  It is commonly chosen to be the surface where the Hubble parameter is constant, and corresponds to its value at the end of inflation \cite{Gibbons:2006pa}, or when it is constant and $\mathcal{O}(\Mp)$ \cite{Kofman:2002cj,Ashtekar:2009mm,Ashtekar:2011rm,Remmen:2013eja,Remmen:2014mia,Sloan:2014jra}.  As discussed in detail in refs.\ \cite{Corichi:2010zp,Schiffrin:2012zf,Corichi:2013kua}, the answer one gets for $P_\text{inf}$ or $\langle \Ntot \rangle$ depends on the choice of $\mathcal{S}$:  For $H \sim \Mp$ one gets $P_\text{inf} \approx 1$ while for $H \ll \Mp$ one gets $P_\text{inf} \ll 1$.

\end{enumerate}

Some of the above listed ambiguities stem from the fact that the underlying Hamiltonian governing the dynamics is constrained to be zero.  We can contrast this to the more familiar situation where one has a physical Hamiltonian $H_\text{p}$ that is non-zero, and there are no remaining constraints on phase space.  In such cases, if we want to calculate probabilities or expectation values we can appeal to ordinary statistical mechanics.  The surface $\mathcal{S}$ referenced in (\ref{eq:P formula}) and (\ref{eq:Ntot formula}) is then naturally replaced by the physical phase space and $d\sigma$ is its natural symplectic measure.  The only quantity to be selected is $\rho$, which can now be interpreted as the phase space density of an ensemble of trajectories evaluated at some ``initial'' time.   That is, $\rho$ defines a probability distribution of initial data.

In  a cosmological model, the shift from constrained Hamiltonian dynamics (which gives  highly ambiguous formulae for characterizing the likelihood of inflation) to unconstrained dynamics (which gives   less ambiguous formulae) is achieved via the time gauge fixing $t=f(a)$ described in Section \ref{sec:time gauges}.   

This procedure leads to the physical 2-dimensional phase space $(\varphi,p_{\varphi}) \in \mathbb{R}^{2}$, and gives the following formula for the expected number of e-folds:
\begin{equation}\label{eq:e-fold final formulae}
	\langle \Ntot \rangle = \frac{{\displaystyle \iint \rho(t_{0},\varphi,p_{\varphi}) \Ntot(t_{0},\varphi,p_{\varphi}) \, d\varphi\,dp_{\varphi}}}{{\displaystyle \iint \rho(t_{0},\varphi,p_{\varphi}) \, d\varphi\,dp_{\varphi}}},
\end{equation}
with a similar formulae for $P_\text{inf}$.  Here, $t_{0}$ is a time at which we specify initial data, $\rho(t_{0},\varphi,p_{\varphi})$ is the phase space distribution describing a statistical ensemble of universes evaluated at $t=t_{0}$, and $\Ntot(t_{0},\varphi,p_{\varphi})$ tells us how initial data at $t=t_{0}$ maps onto the total number of e-folds.  Note that equations (\ref{eq:Ntot formula}) and (\ref{eq:e-fold final formulae}) are essentially identical once we identify $\mathcal{S}$ as the intersection of the gauge-fixing $t_{0}=f(a)$ and Hamiltonian $\mathcal{H}=0$ constraints.  

We can go no further without specifying the phase space density.  We note that in statistical mechanics, $\rho$ evolves as
\begin{equation}
	\frac{\di\rho}{\di t} + \{\rho,H_\text{p}\}=0.
\end{equation}
Systems in thermal equilibrium have $\di\rho/\di t = 0$, which is achieved by taking $\rho$ to be a function of the physical Hamiltonian $H_\text{p}$.  

An interesting assumption is to apply this notion to polymer inflation; that is, we assume that the universe resembles a physical system in thermal equilibrium at early times. Specifically, we assume that during the polymer inflation epoch, the phase space density $\rho$ is a function of one of the physical Hamiltonians discussed in Section \ref{sec:time gauges}.  To specify exactly which Hamiltonian, we can push the thermal analogy a bit further by noting that true equilibrium states are governed by time independent Hamiltonians.  We therefore demand that the physical Hamiltonian in the polymer phase have the same property.  Recalling that the physical Hamiltonian is related to the Hamiltonian density via $H_\text{p} = V_{0}\mathcal{H}_\text{p}$, we see that by enforcing the volume time gauge $t=\Mp^{-1}(a/a_{0})^{3}$, the physical Hamiltonian is
\begin{equation}\label{eq:thermal Hamiltonian}
	 H_\text{p} = 2\Mp^{3} V_{0} a_{0}^{3} H \approx 2\Mp^{4} V_{0} a_{0}^{3} \left(\frac{1}{6} \delta^{2}\varphi^{2} + \frac{1}{12\gamma^{4}} \right)^{1/2},
\end{equation}
which is indeed effectively time independent at early times (i.e., when ${\gamma^{2}}{e^{-3\efolds}} \gg 1$).  

Under these assumptions, the expectation value of the number of (slow roll) e-folds is
\begin{equation}\label{eq:e-fold really final formulae}
	\langle \Ntot \rangle = \frac{{\displaystyle \iint \rho(H_\text{p}) \Ntot(t_{0},\varphi,p_{\varphi}) \, d\varphi\,dp_{\varphi}}}{{\displaystyle \iint  \rho(H_\text{p}) \, d\varphi\,dp_{\varphi}}},
\end{equation}
where $H_\text{p}$ is given by (\ref{eq:thermal Hamiltonian}).  Noting that in the polymer phase, $\varphi$ and $p_{\varphi}$ are essentially constant and equal to their values in the asymptotic past, we can use  (\ref{eq:Ntot 1}) and (\ref{eq:Ntot 2}) to see that $\Ntot \sim \varphi^{2}$ (assuming the quadratic potential).  Hence, the uniform distribution suggested by Laplace's principle $\rho(H_\text{p})=1$ leads to an infinite value of $\langle \Ntot \rangle$.  This is reminiscent of the divergences manifest in the non-gauge fixed approach assuming a flat probability distribution \cite{Gibbons:1986xk,Gibbons:2006pa,Ashtekar:2009mm,Ashtekar:2011rm,Corichi:2010zp,Corichi:2013kua}.  One could attempt to resolve this via some sort of regularization scheme, but this strategy has been criticized \cite{Schiffrin:2012zf}.

\begin{figure}
\begin{center}
\includegraphics[width=\columnwidth]{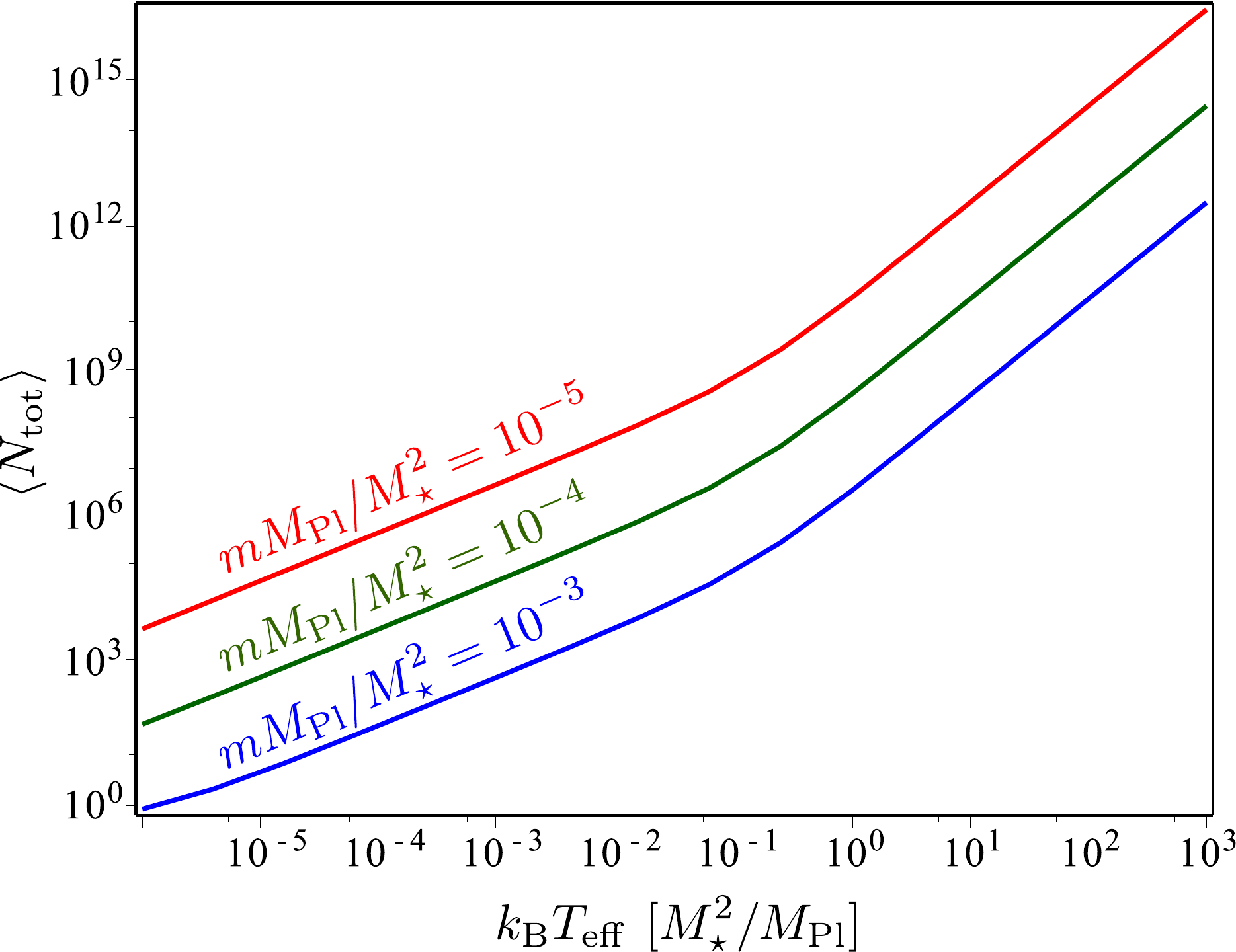}
\end{center}
\caption{Ensemble average of the number of e-folds of slow roll inflation as a function of effective temperature in the polymer phase (assuming a quadratic potential).}\label{fig:N}
\end{figure}
We now demonstrate that one can obtain a finite answer (without resorting to a brute force regularization scheme) for a quadratic potential by assuming a Boltzmann distribution:
\begin{equation}\label{eq:canonical distribution}
	\rho(H_\text{p}) = \exp (-\beta H_\text{p}) = \exp( -H/k_\text{B} T_\text{eff}),
 \end{equation} 
where the effective temperature is
\begin{equation}
	T_\text{eff} = \frac{1}{2k_\text{B} \beta_{0} V_{0} a_{0}^{3} \Mp^{3}}.
\end{equation}
For this potential, the system's dynamics are invariant under $(\varphi,p_{\varphi}) \mapsto (-\varphi,-p_{\varphi})$, so we can restrict the integration in (\ref{eq:e-fold really final formulae}) to $p_{\varphi} \ge 0$.  Making use of $\varphi \approx \varphi_{0}$ and $p_{\varphi} \approx p_{0}$ in the polymer phase, equations (\ref{eq:Ntot 1}) and (\ref{eq:Ntot 2}) give us $\Ntot$ in terms of $\varphi$ and $\text{sgn}(p_{\varphi})$.  But the restricted integration domain implies $\text{sgn}(p_{\varphi}) = 1$, so $\Ntot$ is effectively a function of $\varphi$ only.  Since $H$ is also independent of $p_{0}$, the (infinite) $p_{0}$ integration cancels from the numerator and denominator of (\ref{eq:Ntot formula}).  If we measure $k_\text{B} T_\text{eff}$ in units of $\M^{2}/\Mp$, we see the resulting formula for $\langle \Ntot \rangle$ depends on temperature and $\gamma^{2}\delta = m\Mp/\M^{2}$ only.  The integrals can be computed numerically, and results are shown in Fig.\ \ref{fig:N}.  We see that the expected number of e-folds shows a broken power-law dependence on temperature, with low temperatures associated with a small number of e-folds.  We also see that the number of e-folds increases as $\M$ increases, and decreases as $m$ increases.  

This does lead to the slightly perplexing conclusion that the expected number of e-folds becomes infinite in the $m \rightarrow 0$ limit.  It is not hard to see how this effect arises:  When $m \ll \M$, there is a large hierarchy between the Hubble factor in the polymer and slow-roll phases.  This means that the kinetic-dominated phase lasts a long time.  Since $\phi \propto N$ during this phase, the will generally be a large field amplitude at the start of slow-roll inflation yielding a large number of e-folds.

In closing this section we note that there remains one ambiguity in our approach. This is the choice of time gauge. It is clear that different gauges lead to different physical Hamiltonians. This is turn obviously affects the calculation of $\langle N_\text{tot} \rangle $. The resolution of this issue  would require a solution of the problem of time in quantum gravity, or at least a model that suggests a natural time gauge, such as general relativity coupled to pressureless dust \cite{Husain:2011tk}.

\section{Conclusions}

We have described semiclassical homogeneous cosmological evolution with a scalar field in the context of the polymer quantization method motivated by LQG.   The formalism and calculations are in the framework of a non-vanishing time-dependent physical Hamiltonian obtained by fixing a time gauge. 

The dynamics associated with a quadratic scalar potential unfolds in four distinct phases.  The first is the polymer phase, where the energy density is effectively independent of the field momentum.  This leads to almost exactly de Sitter evolution extending into the infinite past and there is no big bang singularity. This  is followed by a period of kinetic domination where the scalar momentum dominates the potential.  The universe then enters a phase of slow-roll inflation where solutions receive various types of polymer corrections.  The last phase is the usual epoch of reheating where $\phi$ oscillates about the zero of its potential. 

Assuming the polymer energy scale is sufficiently larger than the geometric mean of the scalar and Planck masses, we calculated the effects of polymer quantization on the homogeneous dynamics during slow-roll inflation.  It remains to be seen what the semi-classical polymer effects are on the generation of primordial perturbations.  (An approach to this problem logically distinct from the semi-classical approximation was presented in \cite{Seahra:2012un}.)   We hope to report on this in the future. 

For quadratic potentials, we also demonstrated that it is possible for this model to have significant amount of slow-roll inflation if the initial value of the scalar field amplitude is sub-Planckian.  This is not possible in standard quantizations, where one must assume large initial field values.  We note that the (controversial) BICEP2 results seem to favour models where the scalar amplitude is large at the onset of slow roll inflation \cite{Ade:2014xna}.

We also presented a novel approach for characterizing the probability of inflation using a physical Hamiltonian arising from time gauge fixing.  While we only applied this formalism to the polymer quantized model  presented here, it would be interesting to analyze standard single field inflation and  other models from this perspective.

In a broader context, this work is in the spirit of addressing cosmological evolution in the context of models suggested by possible features of quantum gravity. In our case  this is the low energy consequences of fundamental discreteness as built into the polymer quantization method.  

\begin{acknowledgments}

We would like to thank David Sloan for useful correspondence about the probability of inflation.  We would also like to thank an anonymous referee whose suggestions lead to material improvements to this paper. This work was supported by NSERC of Canada.  S.M.H. was also supported by  the Lewis Graduate Fellowship.
 
\end{acknowledgments}

\bibliography{Massive_Polymer_Inflation}

\end{document}